\documentclass[twocolumn,showpacs,prb,amsmath,amssymb,floatfix,eqsecnum]{revtex4}
\usepackage{amsmath,amssymb}
\usepackage{bm}
\usepackage{epsfig}
\usepackage{psfrag}

\begin{document}

\title{Ferromagnetism in one-dimensional metals: Breakdown of the
Hartree-Fock approximation and  possible first-order phase transition}

\author{Philipp Zedler and Peter Kopietz}
  \affiliation{Institut f\"{u}r Theoretische Physik, Universit\"{a}t
    Frankfurt,  Max-von-Laue Strasse 1, 60438 Frankfurt, Germany}
%
\date{July 26, 2005}
\begin{abstract}
We calculate the Gibbs potential $\Gamma ( M )$ of 
a one-dimensional metal
at constant magnetization $M$
to second order in the screened electron-electron interaction $U$.
At zero temperature we find that $\Gamma (M )$ contains  non-analytic
corrections proportional to $ M^2 \ln | M| $ and $ | M |^3$, 
implying  that a possible 
paramagnetic-ferromagnetic
quantum phase transition in  one-dimensional metals must be first order.

\end{abstract}
\maketitle

\section{Introduction}

Can the ground state of a one-dimensional ($1D$) clean metal exhibit
spontaneous ferromagnetism? A rigorous theorem due to Lieb
and Mattis \cite{Lieb62} implies that the answer to this question is
``no'' for   $1D$
continuum models as well as for $1D$ lattice models with nearest neighbor hopping and
interactions involving only densities. 
However, the Lieb-Mattis theorem does not apply
to  lattice models with longer  range hoppings. 
Indeed, some time ago Daul and Noack \cite{Daul98} presented numerical evidence
that the $1D$ Hubbard model with 
nearest- and next-nearest neighbor hopping 
has a ferromagnetic ground state in a substantial
range of densities and on-site interactions $U$.
Given the stability of the ferromagnetic ground state in $1D$  
in a certain parameter regime, one might want to know the critical 
behavior of the system close to the quantum phase 
transition separating the paramagnetic from the ferromagnetic regime.
Some  physical properties of ferromagnetic metals
in $1D$ have been studied  in several  recent works 
\cite{Bartosch03,Yang04,Sengupta05}.  
However, the fluctuation corrections to the 
Hartree-Fock approximation for the ground state energy
have not been thoroughly investigated.
In view of the fact that the
metallic state in $1D$ is  a Luttinger liquid,
it is not clear whether the Hartree-Fock scenario of a
second order phase transition to a ferromagnetic state for sufficiently strong
interaction is at least qualitatively correct.
In this work we shall therefore calculate the leading (second order in $U$) 
correction
to the Hartree-Fock appoximation for the 
Gibbs-potential $\Gamma ( M )$ at constant magnetization $M$.
We find non-analytic terms which 
completely invalidate the Hartree-Fock prediction
and imply  that, if the  paramagnetic-ferromagnetic quantum 
phase transition exists  in a $1D$ metal, then it  must be first order.
Recently Belitz and Kirkpatrick \cite{Belitz02}
came to a similar conclusion 
about the  paramagnetic-ferromagnetic transition
in  clean itinerant ferromagnets  
in dimensions $1 < D < 3$.

Although most of our
considerations are fairly general and independent of any specific 
model, for explicit calculations we shall use  the 
Hubbard model with nearest neighbor hopping $t$ and next-nearest 
neighbor hopping  $t^{\prime}$ 
on a one-dimensional lattice with $N_L = L/a$ sites and lattice spacing $a$.
The hamiltonian is
 \begin{equation}
 \hat{H} = \sum_{ k \sigma} \epsilon_{ k} \hat{c}^{\dagger}_{ k \sigma} 
 \hat{c}_{ k \sigma} + U \sum_i \hat{n}_{ i \uparrow} \hat{n}_{i \downarrow}
 \; .
 \label{eq:hamiltonian}
 \end{equation}
Here $\hat{c}_{k \sigma} = {N_L}^{-1/2} \sum_i e^{- i k x_i} \hat{c}_{ i \sigma}$ 
annihilates an electron with momentum $k$  and spin $\sigma$,
the operator $\hat{n}_{i \sigma} = \hat{c}^{\dagger}_{i \sigma} 
\hat{c}_{i \sigma}$ counts the number of electrons
with spin $\sigma$ at lattice site $x_i$, and
the energy dispersion is
 \begin{equation}
 \epsilon_{ k } = - 2  t \cos ( k a ) - 2 t^{\prime} \cos ( 2 ka )
 \; .
 \label{eq:dispersion}
 \end{equation}
 
\section{Renormalized perturbation theory at constant magnetization}

\subsection{General considerations}

\begin{figure}[tb]    
   \centering
  \psfrag{lambda}{$\gamma $}
 \psfrag{n}{$n$}
  \vspace{7mm}
      \epsfig{file=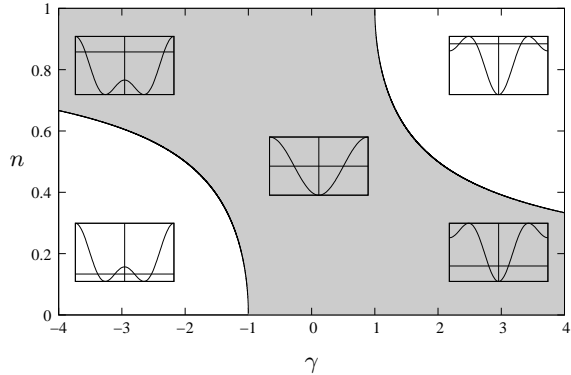,width=75mm}
  \vspace{4mm}
  \caption{%
Topology of the Fermi surface 
for the energy dispersion (\ref{eq:dispersion}) as a function of
$\gamma = 4 t^{\prime} / t $ and filling $n$.
The Fermi surface consists of two points in the shaded regime and has four points in the
white regime. The insets show
the typical dispersions with the position of the chemical potential as a horizontal  line.
}
    \label{fig:dispersion}
  \end{figure}

In the presence of a uniform magnetic field $h$ the grand canonical
potential is 
 \begin{equation}
 \Omega ( \mu , h )   = - T \ln \left\{ 
 {\rm Tr}  \exp [  - \beta ( \hat{H} - \mu \hat{N} - h
 \hat{M} )  ] \right\}
 \; ,
 \label{eq:omegadef}
 \end{equation}
where $\mu$ is the chemical potential, $T = 1 / \beta$ is the
temperature,
$\hat{N} = \sum_{ k \sigma} \hat{c}^{\dagger}_{k \sigma} \hat{c}_{k \sigma}$ 
is the particle number operator, and the operator 
$\hat{M} = \sum_{ k \sigma} \sigma \hat{c}^{\dagger}_{k \sigma} \hat{c}_{k \sigma}$ 
represents the uniform magnetization.
The expectation values of these operators are  
\begin{equation}
 N = \langle \hat{N} \rangle = - \partial \Omega / \partial \mu 
 \; , \; M  = \langle  \hat{M}   \rangle = - \partial \Omega / \partial h
 \; .
 \label{eq:NMdef}
 \end{equation} 
To study spontaneous symmetry breaking, it is more convenient to
work with 
the corresponding Gibbs potential \cite{Nozieres86,Georges91,Kollar03}
\begin{equation}
 \Gamma ( N, M ) = \Omega ( \mu ( N,M), h ( N,M)) + \mu N + h M
 \; ,
 \end{equation}
which is a function of $N$ and $M$. Here
$\mu ( N, M)$ and $h ( N,M)$ should be calculated by inverting
Eqs.~(\ref{eq:NMdef}).
Perturbation theory generates an expansion of
$\Gamma ( N , M )$ in powers of the relevant 
dimensionless interaction, the so-called  Stoner factor\cite{Moriya85}
 \begin{equation}
 I = U \nu_0 
 \; ,
 \end{equation}
where $\nu_0 = \nu ( \xi =0) $ is the density of states (DOS)
per spin projection  at the Fermi energy of the non-interacting system 
in the absence of a magnetic field. 
We define the energy-dependent DOS via  
\begin{equation}
 \nu ( \xi )  = \frac{1}{N_L} \sum_{ k } \delta ( \xi - \epsilon_k + \mu )
 \; ,
 \label{eq:nuxidef}
 \end{equation}
which has units of inverse energy.
It is convenient to
measure particle number and magnetization in reduced
units
 \begin{equation}
 n = \frac{N}{2N_L} \; , \; m = \frac{M}{2 N_L}
 \, ,
 \label{eq:nmdef}
 \end{equation}
so that  $ 0 \leq n \leq 1$ and $ - \frac{1}{2} \leq m \leq \frac{1}{2}$.
Note that the physical range of $m$-values
is $ | m | \leq n$ for $ n \leq 1/2$ and $| m | \leq 1-n$ for $ n \geq 1/2$, because the
magnetization $M$  cannot exceed the number $N$ of electrons
for a less than half-filled band or the number 
$2 N_L - N$ of holes for a more than half filled band.
Depending on the values of the two dimensionless parameters $n$ 
and 
 \begin{equation}
 \gamma = 4 t^{\prime} / t
 \; ,
 \label{eq:alphadef} 
 \end{equation}
the Fermi surface of our model without interaction and magnetic field
consists  of two or four discrete points. The different regimes are
shown in Fig.~\ref{fig:dispersion}.
(A similar figure can be found in Ref.~[\onlinecite{Daul98}].)
For $ | \gamma | < 1$ the Fermi surface has two Fermi points for all
fillings $ 0 < n < 1$.  For $ \gamma > 1$, i.e. for $ t^{\prime} > t/4$,
the Fermi surface has two points for $ 0 < n < n_c ( \gamma )$, and
four points for $ n_c ( \gamma ) < n < 1$. The critical filling separating these regimes
is
 \begin{equation}
 n_c ( \gamma ) = \frac{1}{\pi} \arccos \left( 1 - \frac{2 }{ \gamma} \right)
 \; , \; \gamma > 1 
 \; .
 \label{eq:ncdef}
 \end{equation}
On the other hand, for $\gamma < -1$, corresponding to $t^{\prime} < - t / 4$,
the Fermi surface has four points at low fillings $ 0 < n < n_c ( \gamma )$,
and two points at larger fillings $ n_c ( \gamma ) < n < 1$, where now
 \begin{equation}
 n_c ( \gamma ) = \frac{1}{\pi} \arccos \left( \frac{2}{| \gamma | } -1 \right)
 \; , \; \gamma < -1 \; .
 \label{eq:ncdef2}
 \end{equation}
Besides the usual singularities for $ n \rightarrow 0$ 
and $ n \rightarrow 1$, for $ | \gamma | > 1$
the DOS at the Fermi energy exhibits additional
one-sided singularities at the critical 
fillings $n_c ( \gamma )$, which are related to vanishing
Fermi velocities when the Fermi surface topology changes discontinuously.

\begin{figure}[tb]
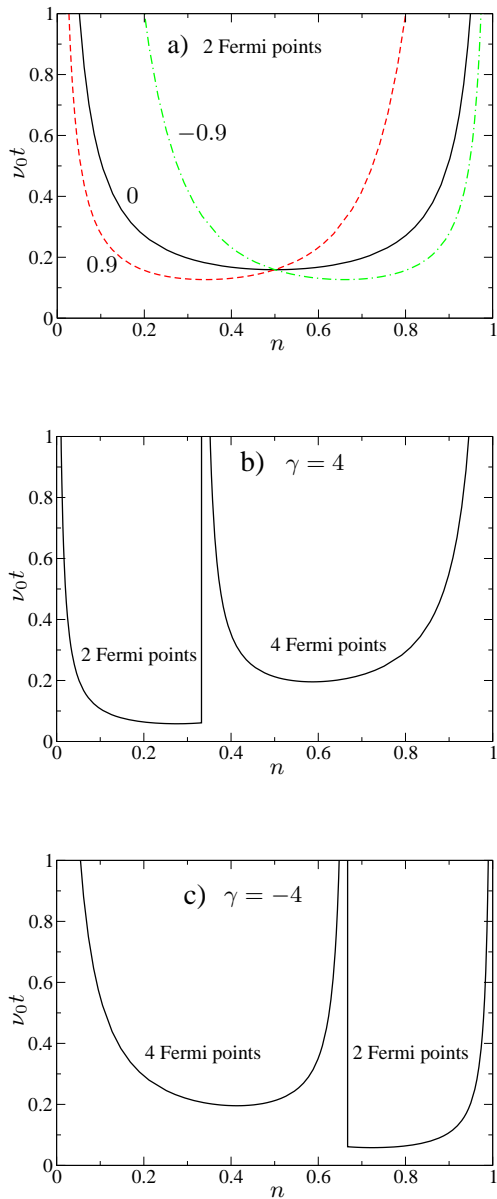
    
   \centering
  \psfrag{dos}{$ \nu_0 t$}
 \psfrag{n}{$n$}
 \psfrag{g0}{$0$}
 \psfrag{g9}{$ 0.9$}
 \psfrag{g4}{$\gamma = 4$}
 \psfrag{g4m}{$\gamma =-4$}
 \psfrag{gm9}{$ -0.9$}
 \epsfig{file=fig2a.eps,width=65mm}
\vspace{10mm}

 \epsfig{file=fig2b.eps,width=65mm}
\vspace{10mm}

\epsfig{file=fig2c.eps,width=65mm}
  \vspace{7mm}
  \caption{%
(Color online) Density of states at the Fermi energy
in units of $t^{-1}$ 
for the dispersion (\ref{eq:dispersion})
as a function of filling $n$ for 
different values of $\gamma = 4 t^{\prime} /t$, see
Eqs.(\ref{eq:dos2fp}, \ref{eq:dos4fpa}, \ref{eq:dos4fpb}). 
The numbers close to the curves in a) 
give the corresponding values of $\gamma$.
Note that for $\gamma > 1$ the DOS remains finite
for $n \rightarrow n_c$ from below, but
diverges if $n$ approaches $n_c$ from above.
In contrast, for $ \gamma < -1$ the DOS 
is finite if $n \rightarrow n_c$ from above and is
singular if $n \rightarrow n_c$ from below.
}
    \label{fig:dos}
  \end{figure}

In the regime where the Fermi surface consists of two points
the DOS at the Fermi energy is 
 \begin{equation}
 \nu_0 = \frac{a}{ \pi v_F} = \frac{1}{ 2 \pi t \sin ( \pi n ) [ 1 + \gamma \cos ( \pi n ) ] }
 \; ,
 \label{eq:dos2fp}
 \end{equation}
where $v_F$ is the Fermi velocity.
The DOS 
is more complicated in the regime where the Fermi surface has four points.
For $\gamma > 1$ and $ n > n_c$ we
find
 \begin{eqnarray}
 \nu_0  & = &  \frac{2}{\pi t \gamma}
 \left[
 \frac{1}{ \tan ( \pi (1 -n )/2)} \right]
 \nonumber
 \\
 & & \times
 \left[
 \frac{    1 + \cos( \pi (1-n ))   }{ [ 1 + \cos( \pi (1-n ) ) ]^2 -  4 / \gamma^2 }
 \right]
 \; ,
 \label{eq:dos4fpa}
 \end{eqnarray}
and for $ \gamma < -1$ and $n < n_c$,
\begin{eqnarray}
 \nu_0  & =  & \frac{2}{\pi t | \gamma |}
 \left[
 \frac{1}{ \tan ( \pi n /2)} \right]
 \nonumber
 \\
 & & \times
 \left[
 \frac{    1 + \cos( \pi n )   }{ [ 1 + \cos( \pi n  ) ]^2 -  4 / \gamma^2 }
 \right]
 \; .
 \label{eq:dos4fpb}
 \end{eqnarray}
Graphs of the  DOS at the Fermi energy
for different values of $\gamma$
are shown in Fig.~\ref{fig:dos}.
For simplicity we 
assume in this work that in the absence 
of a magnetic field and ferromagnetic symmetry breaking the true
Fermi surface consists  only two points
$ k_F$ and $- k_F$, each of which is split
by the  magnetic field $h$ (or for finite $m$) into two spin-dependent points
$ k_{\sigma}$ and $- k_{\sigma}$, where $ \sigma = \uparrow, \downarrow$.
The calculations in this work are therefore restricted to the
shaded regime shown in Fig.~\ref{fig:dispersion}.
We emphasize that $k_{\sigma}$ 
are the true  Fermi momenta of the 
interacting system, 
which have to be determined self-consistently \cite{Nozieres64,Neumayr03}.
In terms of the dimensionless filling $n$ and the magnetization $m$
defined in Eq.~(\ref{eq:nmdef}) we may write
 \begin{subequations}
 \begin{eqnarray}
 k_{\uparrow} + k_{\downarrow} & = & 2 \pi n /a = \pi N / L
 \; , 
 \\
 k_{\uparrow} - k_{\downarrow} & = & 2 \pi m /a = \pi M / L
 \; .
 \end{eqnarray}
 \end{subequations}
To study ferromagnetic symmetry breaking, we consider the 
change of the  Gibbs potential
due to a finite value of $M$.
For convenience we
introduce the dimensionless magnetization-dependent part of
the Gibbs potential
 \begin{eqnarray}
  g  ( m  ) 
& = &  \frac{ \nu_0}{N_L} 
 \left[ \Gamma ( n , m ) -  \Gamma ( n , m = 0) \right]
 \nonumber
 \\
 & = &   g_0 ( m ) + I   g_1 ( m ) + 
 \frac{ I^2}{2}   g_2 ( m ) + \ldots
 \; .
 \label{eq:gmdef}
 \end{eqnarray}
We suppress the dependence on
the filling factor $n$, which is implicit in all quantities such as
$v_F$,   $\nu_0$ and $I$.
In Eq.~(\ref{eq:gmdef}) we have normalized the Gibbs potential by the natural scale $
 \frac{N_L}{  \nu_0 }  = 
\frac{Lv_F}{\pi} \left( \frac{ \pi}{a } \right)^2$
of the kinetic energy.
In the regime where the Fermi surface without symmetry 
breaking has only two points \cite{footnote1} it is then sufficient to 
set up the renormalized perturbation theory 
by introducing two counterterms
quadratic in the fermions with coefficients
$\Delta_{\uparrow}$ and $\Delta_{\downarrow}$ and rewrite
the operator in the exponential of
Eq.~(\ref{eq:omegadef}) as follows,
 \begin{equation}
 \hat{ H} - \mu \hat{N} - h \hat{M} = \hat{K}_0 + \hat{V}
 \; ,
 \end{equation}
where
 \begin{equation}
 \hat{K}_0 = \sum_{ k \sigma}  \xi_{ k \sigma} 
\hat{c}^{\dagger}_{ k \sigma} 
 \hat{c}_{ k \sigma}
 \; , 
  \end{equation}
and
 \begin{equation}
 \hat{V} = 
U \sum_i \hat{n}_{ i \uparrow} \hat{n}_{i \downarrow}
 - \sum_{ k \sigma} \Delta_{\sigma} 
 \hat{c}^{\dagger}_{ k \sigma} 
 \hat{c}_{ k \sigma} 
\label{eq:Vdef}
 \; ,
 \end{equation}
with
 \begin{equation}
 \xi_{ k \sigma} = \epsilon_k + \Delta_{\sigma} - \mu - \sigma h
 \; .
  \end{equation}
In the language of many-body theory, $\Delta_{\sigma} = 
\Sigma_{\sigma} ( k_{\sigma} , i0 ) $
is the (a priori unknown)
self-energy due to the two-body interaction
of  
our original hamiltonian (\ref{eq:hamiltonian})
for momenta $k = k_{\sigma}$ at the Fermi surface and
for vanishing frequency.
Following the usual procedure \cite{Neumayr03}, 
the counterterms $\Delta_{\sigma}$ can be determined order by order in
perturbation theory by demanding that the
self-energy
generated by the subtracted interaction $\hat{V}$ in Eq.~(\ref{eq:Vdef})
vanishes for $k = k_{\sigma}$ and $\omega =0$.
Alternatively, the self-consistent 
determination of the counterterms can also be
implemented non-perturbatively within the
framework of the renormalization group \cite{Kopietz01,Ledowski05}.

\subsection{Non-interacting limit}

\begin{figure}[tb]    
   \centering
  \psfrag{CCCCC}{${C} $}
 \psfrag{gamma}{ $\gamma$}
 \psfrag{n}{$n$}
  \vspace{7mm}
 \epsfig{file=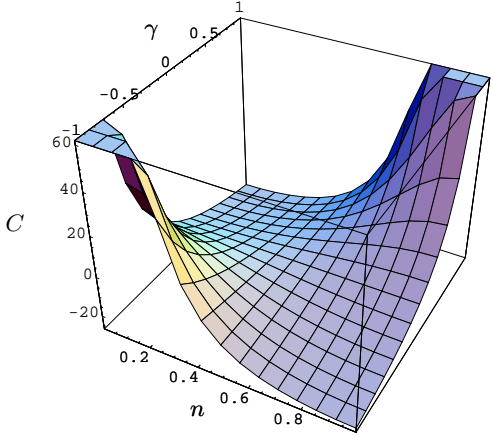,width=65mm}
  \vspace{4mm}
  \caption{%
(Color online) Dimensionless coefficient $C$ defined in Eq.~(\ref{eq:Cdef})
for the energy dispersion given in Eq.~(\ref{eq:dispersion})
in the regime $-1 < \gamma < 1$ and for arbitrary
filling $n$. 
}
    \label{fig:Cplot}
  \end{figure}

\begin{figure}[tb]    
   \centering
  \psfrag{C}{${C} $}
 \psfrag{g9}{ $\gamma=0.9$}
 \psfrag{g9m}{$\gamma=-0.9$}
 \psfrag{g0}{$\gamma =0$}
 \psfrag{n}{$n$}
  \vspace{7mm}
 \epsfig{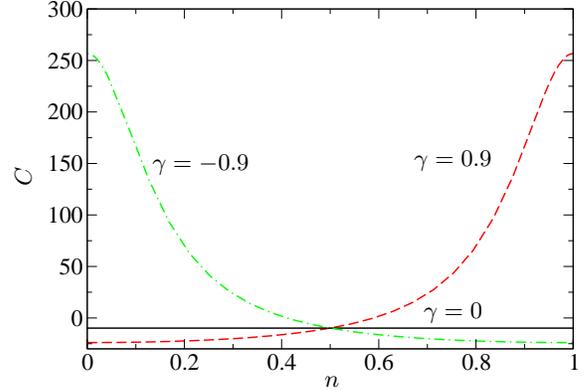}
  \vspace{4mm}
  \caption{%
(Color online) Dimensionless coefficient $C$ defined in Eq.~(\ref{eq:Cdef})
for different values of $\gamma = 4 t^{\prime}/t$ as a function of the filling $n$.
}
    \label{fig:Csection}
  \end{figure}

Let us first consider the
non-interacting limit $U=0$,
where
$\Delta_{\sigma} =0$ and 
 \begin{equation}
 \Omega_0 ( \mu , h ) = - T \sum_{ k \sigma} \ln [ 1 - e^{ - \beta ( \epsilon_k - \mu - \sigma h ) } ]
 \; .
 \end{equation}
The expressions relating particle number and magnetization 
to chemical potential and magnetic field are then
 \begin{subequations}
 \begin{eqnarray}
 N & = & \sum_{ k \sigma } f (  \epsilon_k - \mu - \sigma h ) 
 \label{eq:N0relation}
 \; ,
 \\
M & = & \sum_{ k \sigma } \sigma f (  \epsilon_k - \mu - \sigma h ) 
 \label{eq:M0relation}
 \; ,
 \end{eqnarray}
 \end{subequations}
where $ f ( \epsilon ) = [ e^{ \beta \epsilon } + 1 ]^{-1}$ is the Fermi function.
We denote the corresponding Gibbs potential
by $\Gamma_0 ( N , M )$. In general it is not possible to calculate
$\Gamma_0 ( N , M )$ analytically. However, 
for weak ferromagnets, where $ | m | \ll 1$, 
we may expand $\Gamma ( N , M)$ in powers of $M$.
For the
$m$-dependent part of the
dimensionless Gibbs potential defined in Eq.~(\ref{eq:gmdef})
we obtain in the non-interacting limit at zero temperature,
\begin{equation}
  g_0 ( m ) = m^2 + \frac{C}{12} m^4 + O ( m^6 )
 \; ,
 \end{equation}
where the dimensionless coefficient $C$ of the quartic term
can be written in terms of the derivatives of the DOS
at the Fermi energy \cite{Moriya85,Nozieres86},
 \begin{equation}
 C = \frac{1}{\nu_0^2} \left[ 3 \left( \frac{ \nu^{\prime}_0 }{\nu_0 } \right)^2 -
 \frac{ \nu^{\prime \prime}_0}{\nu_0} \right]
 \label{eq:Cdef}
 \; .
 \end{equation}
Here $\nu_0^{\prime}$ and $\nu_0^{\prime \prime}$ are the first and the second 
derivative of the energy-dependent DOS $\nu ( \xi ) $ 
defined in Eq.~(\ref{eq:nuxidef}) at $\xi =0$.
Equation (\ref{eq:Cdef}) is valid in any dimension provided we use the
$D$-dimensional DOS. Interestingly, if  in $1D$
the Fermi surface without magnetic field has only two points $ \pm k_F$,
then the constant $C$
can be related 
to the cubic term in the expansion of the
energy dispersion around the Fermi surface as follows
 \begin{equation}
 C = \frac{ c_3}{ v_F} \left( \frac{\pi}{a} \right)^2
 \; ,
 \label{eq:Cdef2pf}
 \end{equation}
where $c_3$ is defined by
 \begin{equation} 
 \epsilon_{ k_F + q } = \epsilon_{k_F} + v_F q + \frac{ q^2}{ 2 m^{\ast}} 
 + \frac{c_3}{6} q^3 + O ( q^4 )
 \; .
 \label{eq:epsexpansion}
 \end{equation}
The quadratic term in the expansion (\ref{eq:epsexpansion})
cancels on the right-hand side of Eq.~(\ref{eq:Cdef}).
For the dispersion (\ref{eq:dispersion}) we obtain in the regime
where the Fermi surface consists of two points,
 \begin{equation}
  C = - \pi^2 \frac{ 1 + 4 \gamma \cos ( \pi n )}{ 1 + \gamma \cos ( \pi n )}
 \; .
 \end{equation}
A graph of this expression is shown 
in Fig.~\ref{fig:Cplot}.
For clarity, in Fig.~\ref{fig:Csection} we also show
some cuts for fixed 
$\gamma$ through  the  surface in Fig.~\ref{fig:Cplot}.
While for  $| \gamma | < 1/4$ the coefficient $C$ is negative for
all densities, for $ | \gamma | > 1/4$ there exists always a regime of densities
where $C$ is positive. In fact, by choosing $n$ sufficiently close to $ n_c$ 
we can fine tune
$C$ to assume  any desired positive value.

\subsection{Hartree-Fock approximation}
\begin{figure}[tb]    
   \centering
  \psfrag{m}{$m$}
  \psfrag{gmm}{$ g(m) \times 10^4 $}
  \vspace{7mm}
  \epsfig{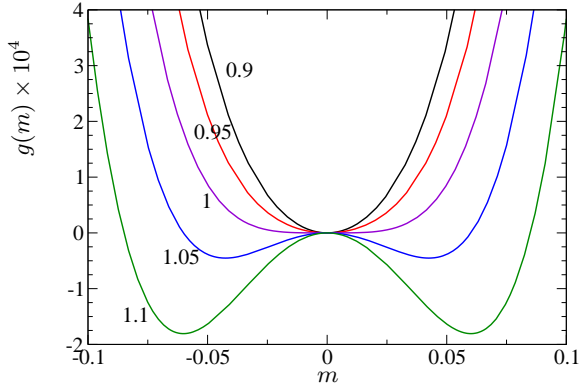}
  \vspace{4mm}
  \caption{%
(Color online) Hartree-Fock approximation for the dimensionless 
Gibbs potential $g ( m )$ given in Eq.~(\ref{eq:gmHF}) for 
$\gamma =0.9$ and $n=0.9$, corresponding to $C \approx 166$.
The numbers close to the curves are the corresponding values of
the Stoner factor $I$.
Note that for $n =0.9$ the physical values for $m$ are in the range
$ | m | \leq 1- n =  0.1$.
}
    \label{fig:Gamma1}
  \end{figure}

To first order in $U$ the self-energy
generated by  the subtracted interaction
$\hat{V}$ is
 $
 \tilde{\Sigma}_{\sigma}  = U n_{ - \sigma} - \Delta_{\sigma}
 $,
which is independent of momentum $k$ and frequency $\omega$.  Here
 \begin{equation}
 n_{\sigma} = \frac{1}{N_L} \sum_{k} f( \xi_{ k \sigma}  )
 \; .
 \end{equation}
From the requirement that 
the subtracted interaction does not generate a momentum- and 
frequency-independent self-energy,
$\tilde{\Sigma}_{\sigma} =0$, we find
the counterterms to first order in $U$,
 \begin{equation}
 \Delta_{\sigma} = U n_{ - \sigma} \equiv \Delta + \sigma \tilde{\Delta}
 \; ,
 \end{equation}
where we have defined
 \begin{subequations}
 \begin{eqnarray}
 \Delta & = & \frac{ \Delta_{\uparrow} + \Delta_{\downarrow}}{2} = U \frac{ 
n_{\uparrow} + n_{\downarrow}}{2} = U n
 \; ,
 \\
 \tilde{\Delta} & = & \frac{ \Delta_{\uparrow} - \Delta_{\downarrow}}{2} = 
U \frac{ 
n_{\uparrow} - n_{\downarrow}}{2} =
U m
 \; .
 \end{eqnarray}
 \end{subequations}
The grand canonical potential is in this approximation given by
 \begin{equation}
 \Omega ( \mu , h )  \approx  \Omega_0 ( \mu - \Delta , h + \tilde{\Delta} )
 - U ( n^2 - m^2 ) N_L
 \; ,
 \end{equation}
and the corresponding Gibbs potential is
 \begin{equation}
 \Gamma ( N , M )  \approx  \Gamma_0 ( N , M ) + U ( n^2 - m^2 ) N_L
 \label{eq:Gibbsres}
 \; .
 \end{equation}
The function $ g_1 ( m )$  defined in Eq.~(\ref{eq:gmdef}) is therefore
  $g_1 ( m ) = - m^2$
so that to first order in $I$ we  obtain
 \begin{equation}
 g ( m ) = ( 1 - I ) m^2 + \frac{C}{12} m^4 + O ( m^6 , I^2)
 \label{eq:gmHF}
 \end{equation}
In the regime where $C > 0$ this leads to the usual Hartree-Fock
scenario \cite{Nozieres86,Moriya85}
of a second order quantum phase transition to a ferromagnetic state at
$I_c  = 1$, as shown in Fig.~\ref{fig:Gamma1}.
Of course, extrapolating the Hartree-Fock approximation to $I \approx 1$
is an uncontrolled procedure, because in this 
interaction range
there is no reason why the
corrections of order $I^2$ and higher in Eq.~(\ref{eq:gmHF}) 
should be small. 
In order to assess the validity of the Hartree-Fock scenario, we 
shall calculate in the following section the correction  to the
Gibbs potential to second order in $I$.

\section{Gibbs potential to second order in the interaction}
\begin{figure}[tb]    
   \centering
  \psfrag{u}{{\huge{$\uparrow$}}}
  \psfrag{d}{{\huge{$\downarrow$}}}
  \vspace{7mm}
     \epsfig{file=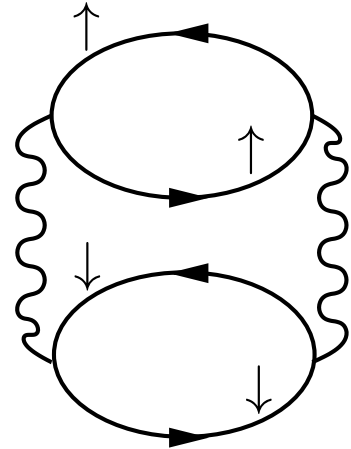,width=45mm}
  \vspace{4mm}
  \caption{%
Feynman diagram representing the correction to the
Gibbs potential to second order in the on-site interaction.
The wavy lines represent the bare interaction and
solid arrows represent the Hartree-Fock Green-functions 
with the indicated spin projections.
}
    \label{fig:Feynman}
  \end{figure}

\begin{figure}[tb]    
   \centering
  \psfrag{gm}{$g (m) \times 10^4 $}
  \psfrag{m}{$m$}
  \vspace{7mm}
      \epsfig{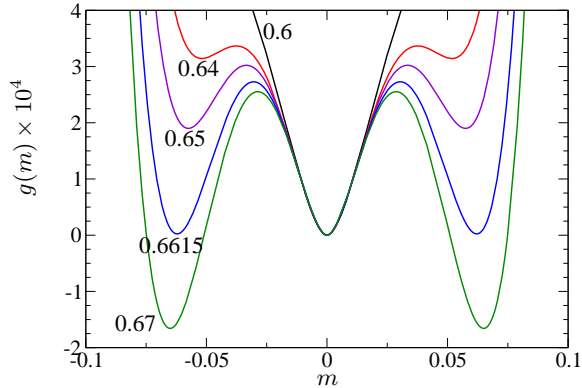}
  \vspace{4mm}
  \caption{%
(Color online) Gibbs potential $g ( m )$ 
to second order in the interaction given in Eq.~(\ref{eq:gm2}) with
$\lambda = (1-n)/ \pi$.
The values of the Stoner factor are written next to the curves.
The parameters
$\gamma =0.9$ and $n=0.9$ as well as the scales
are the same as in Fig.~\ref{fig:Gamma1}.}
    \label{fig:Gamma2}
  \end{figure}

The second order 
correction to the Gibbs potential is given by the
Feynman diagram in Fig.~\ref{fig:Feynman}.
In real space and imaginary time this diagram represents the following
expression,
 \begin{eqnarray}
  \Gamma_2 (N, M ) & = &
 - N_L \frac{U^2  a^4}{2} \int_{- \beta/2}^{\beta /2} d \tau 
  \nonumber
 \\
 &  \times &
\sum_i
 \Pi_{ \uparrow  } ( x_i , \tau )
 \Pi_{ \downarrow  } ( x_i , \tau )
 \; ,
 \label{eq:Gibbs2}
 \end{eqnarray}
where
 \begin{equation}
\Pi_{ \sigma  } ( x , \tau ) = - G_{\sigma} ( x , \tau ) 
 G_{\sigma } (- x , - \tau )
 \end{equation}
is the polarization bubble 
for spin $\sigma$ electrons without interactions. Here
the real-space imaginary-time Green function is
 \begin{equation}
 G_{\sigma} ( x , \tau ) = \frac{ 1}{   \beta L } \sum_{k , \omega_l } 
 \frac{e^{ i ( k x - \omega_l \tau )}}{ i \omega_l - \xi_{ k \sigma}}
 \; ,
 \label{ew:Gdef}
 \end{equation} 
where $\omega_l = 2 \pi ( l + \frac{1}{2} ) T$, $l =0, \pm 1 , \pm 2, \ldots$ 
are fermionic Matsubara
frequencies.
Carrying out the Matsubara sum
and taking the zero temperature limit ($\beta \rightarrow \infty$) and
the infinite system limit ($ L \rightarrow \infty$)  we obtain
\begin{equation}
 G_{\sigma} ( x , \tau ) = \int_{ - \pi /a }^{\pi / a } \frac{ d k}{ 2 \pi}
  e^{i kx } G_{k \sigma } ( \tau )
 \; ,
 \label{eq:Gxtau2}
 \end{equation}
with
 \begin{equation}
  G_{k \sigma } ( \tau ) = - e^{ - \xi_{ k \sigma} \tau }
 \left[ \Theta ( \tau ) \Theta ( \xi_{ k \sigma} ) -
 \Theta ( - \tau ) \Theta ( - \xi_{ k \sigma} ) \right]
 \; .
 \end{equation}
To make progress analytically, we
linearize the energy dispersion
within an interval $ - \Lambda < q < \Lambda$ around the Fermi points.
Here $\Lambda$ is an ultraviolet cutoff of the order of $1/a$ and we assume
that 
\begin{equation}
 m \lesssim  \frac{ \Lambda a }{ \pi } \equiv  \lambda \; ,
 \label{eq:assumption}
 \end{equation}
which is necessary in order to justify the linearisation of the
energy dispersion at the Fermi points. 
The dominant correction term to the Hartree-Fock approximation 
turns out to depend only logarithmically on $\Lambda$, so that
it is not very sensitive to the numerical value of $\Lambda$.
With these assumptions
the integration in Eq.~(\ref{eq:Gxtau2})
can be carried out exactly, 
 \begin{eqnarray}
 G_{\sigma} ( x , \tau )
 & \approx & \frac{1}{ 2 \pi i } \Bigl[
 e^{i k_{\sigma  } x} 
 \frac{ 1 - e^{ i \Lambda s_{ \tau}  ( x  + i v_{\sigma} \tau ) } }{
 x + i v_{\sigma} \tau }
\nonumber
 \\
 & & - 
 e^{- i k_{\sigma  } x } 
 \frac{ 1 - e^{ - i \Lambda s_{ \tau}  ( x  - i v_{\sigma} \tau ) } }{
 x - i v_{\sigma} \tau }
 \Bigr]
 \; ,
 \label{eq:Gdef}
 \end{eqnarray}
where $s_{\tau} = {\rm sign} ( \tau )$
and
 $v_{\sigma} = \partial \xi_{ k \sigma} / {\partial k} |_{ k = k_{\sigma}}$.
In this approximation the polarization can be written as
 \begin{equation}
 \Pi_{\sigma } ( x , \tau ) =
\Pi_{\sigma }^0 ( x , \tau ) 
 +
\Pi_{\sigma }^{2 k_\sigma}  ( x , \tau )
 +
\Pi_{\sigma }^{- 2 k_\sigma}  ( x , \tau )
 \label{eq:Pisplit}
 \; ,
 \end{equation}
with the forward scattering contribution
 \begin{equation}
\Pi_{\sigma }^0 ( x , \tau )  = 
- \frac{2}{ ( 2 \pi )^2} {\rm Re} \left[ 
 \frac{ 1 - e^{ i \Lambda s_{\tau} ( x + i v_{\sigma} \tau ) } }{ x + i v_{\sigma} \tau } \right]^2
 \; ,
 \end{equation}  
and the backscattering part
 \begin{equation}
\Pi_{\sigma }^{2 k_{\sigma} } ( x , \tau )  = 
 \frac{  e^{2 i k_{\sigma} x} }{ ( 2 \pi )^2} 
 \frac{  |  1 - e^{ i \Lambda s_{\tau} ( x + i v_{\sigma} \tau ) } |^2  }{ 
 x^2 +  ( v_{\sigma} \tau )^2 } 
 \; .
 \label{eq:Piback}
 \end{equation}  
Substituting Eq.~(\ref{eq:Gdef}) into Eq.~(\ref{eq:Gibbs2}) and taking the limts
$\beta \rightarrow \infty$ and $ L \rightarrow \infty$, we find
\begin{widetext}
 \begin{eqnarray}
  \Gamma_2 ( N , M ) & = & - N_L U^2 a^3 \frac{4}{ ( 2 \pi )^4 }
 \int_{ 0 }^{\infty} d \tau \int_{ - \infty}^{\infty} dx
 \nonumber
 \\
 & \times &  \Biggl\{
 {\rm Re} \left[
 \frac{ 1 - 2 g_{\uparrow} e^{ i \Lambda x } + g_{\uparrow}^2 e^{ 2 i \Lambda x}}{
 ( x + i v_{\uparrow} \tau )^2 }
 \right]
 {\rm Re} \left[
 \frac{ 1 - 2 g_{\downarrow} e^{ i \Lambda x } + g_{\downarrow}^2 e^{ 2 i \Lambda x}}{
 ( x + i v_{\downarrow} \tau )^2 }
 \right]
 \nonumber
 \\
 & & -
 \cos ( 2 k_{\uparrow} x ) \left[ \frac{ 1 - 2 g_{\uparrow} \cos ( \Lambda x ) + g_{\uparrow}^2 }{
 x^2 + ( v_{\uparrow} \tau )^2 } \right]
 {\rm Re} \left[
 \frac{ 1 - 2 g_{\downarrow} e^{ i \Lambda x } + g_{\downarrow}^2 e^{ 2 i \Lambda x}}{
 ( x + i v_{\downarrow} \tau )^2 }
 \right]
\nonumber
 \\
 & & -
 \cos ( 2 k_{\downarrow} x ) \left[ \frac{ 1 - 2 g_{\downarrow} \cos ( \Lambda x ) + g_{\downarrow}^2 }{
 x^2 + ( v_{\downarrow} \tau )^2 } \right]
 {\rm Re} \left[
 \frac{ 1 - 2 g_{\uparrow} e^{ i \Lambda x } + g_{\uparrow}^2 e^{ 2 i \Lambda x}}{
 ( x + i v_{\uparrow} \tau )^2 }
 \right]
 \nonumber
 \\
 & & +
 \cos ( 2 k_{\uparrow} x ) \cos ( 2 k_{\downarrow} x )
\left[ \frac{ 1 - 2 g_{\uparrow} \cos ( \Lambda x ) + g_{\uparrow}^2 }{
 x^2 + ( v_{\uparrow} \tau )^2 } \right]
 \left[
\frac{ 1 - 2 g_{\downarrow} \cos ( \Lambda x ) + g_{\downarrow}^2 }{
 x^2 + ( v_{\downarrow} \tau )^2 } \right]
 \Biggr\}
 \; .
 \label{eq:GG2}
 \end{eqnarray}
\end{widetext}
Here $ g_{\sigma} = e^{ - v_{\sigma} \Lambda \tau }$.
The $x$-integration can now be performed using the residue theorem.
The resulting $\tau$-integration can then  be carried out  exactly.
Keeping in mind that the difference $v_{\uparrow} - v_{\downarrow}$ between the
Fermi velocities is proportional to $m U$, 
we may approximate
$v_{\uparrow} \approx v_{\downarrow} \approx v_F$ in Eq.~(\ref{eq:GG2}), because
the prefactor is already of order $U^2$.
In this approximation we obtain
 \begin{equation}
  g_2 (m ) \equiv  \frac{ \nu_0}{N_L} 
 \left[ \Gamma_2 ( N , M ) - \Gamma_2 ( N , 0 ) \right] =
\lambda^2 f ( m  / \lambda   )
 \; ,
 \label{eq:g2res} 
\end{equation}
with
 \begin{eqnarray}
 f ( x ) & = & - x^2 \ln | x | - 3 | x | - | x | ( 1 -2 |x | )  \ln ( 1 - 2 |x | )
 \nonumber
 \\
 &  & - (1-x^2 ) \ln ( 1 -| x | )
 \nonumber
 \\
 &  & + ( 1 + |x | - 2 x^2 ) \ln ( 1 + 2 | x | )
 \; .
 \label{eq:tildeg2res}
 \end{eqnarray} 
Because  Eqs.~(\ref{eq:g2res}) and (\ref{eq:tildeg2res}) 
have been derived assuming 
that $x = m / \lambda$ is small (see Eq.~(\ref{eq:assumption})) 
it is
consistent to expand $f (x )$ in powers of $x$,
 \begin{equation}
 f ( x ) =   \left[ \frac{5}{2}    - \ln|x|  \right] x^2  - 6 |x|^3 + \frac{13}{12} x^4 + 
O ( x^5) \; .
 \end{equation} 
Then we obtain the magnetization-dependent part of the
normalized Gibbs potential to second order in $I$ and to fourth order in $m$, 
 \begin{eqnarray}
  g ( m ) & = &  \left\{ 1 - I + \frac{I^2}{2} 
 \left[    \ln \left( \frac{\lambda}{ | m | } \right) 
 +  \frac{5}{2}     \right] \right\} 
m^2  - \frac{3 I^2}{\lambda} |m| ^3 
 \nonumber
 \\
 &  & +
  \left( C + \frac{13 I^2 }{2 \lambda^2 }   \right) \frac{m^4}{12} + O ( | m |^5, I^3 )
\; .
\label{eq:gm2}
 \end{eqnarray}
A reasonable choice for the
ultraviolet cutoff
 $\Lambda$ defining the interval where the
linearization of the energy dispersion is justified 
is $\Lambda = n/a$ for $ n \leq 1/2$ and
$\Lambda = (1-n ) / a$ for $ n > 1/2$.  
Our dimensionless cutoff $\lambda$ 
defined in Eq.~(\ref{eq:assumption}) 
is then  $\lambda =  \pi^{-1} {\rm min} \{ n, 1-n \}$. 
A graph of Eq.~(\ref{eq:gm2}) for this choice of 
$\lambda$  and $n=0.9$  is shown in Fig.~\ref{fig:Gamma2}.
The non-analytic terms proportional to $m^2 \ln | m | $ and $ | m |^3$
generated by the second order correction completely
change  the
Hartree-Fock scenario of a second order quantum phase transition
depicted in Fig.~\ref{fig:Gamma1}.  From Fig.~\ref{fig:Gamma2} it is obvious
that the extrapolation of the second order correction to an
interaction  strength of the order of unity
leads to a  first order
quantum phase transition, which for
our choice of $\gamma = 0.9$ and filling $n =0.9$ occurs
 at a critical value  
$ I_c \approx 0.6615$.
At this value of $I$ our function
$g ( m )$ develops three separate degenerate minima,
which is the characteristic feature of a first order phase transition.
The precise numerical  value for the critical $I_c$
depends on our particular choice of the cutoff $\lambda$, so that
the above value of $I_c$ should not be taken too serious.  
However, with any reasonable choice
of $\lambda$ the critical $I_c$ is smaller than the  
Hartree-Fock result $I_c =1$, suggesting that
correlations  can stabilize the ferromagnetic state
in certain parameter regimes.
For consistency, we should require that the magnetization
$m$ at the phase transition satisfies $m \lesssim \lambda \approx (1-n) / \pi$, see
Eq.~(\ref{eq:assumption}). From Fig.~\ref{fig:Gamma2} we see that this
condition is only marginally satisfied, so that
our approximations loose their
quantitative accuracy once the curves in  Fig.~\ref{fig:Gamma2}
develop minima at finite  $m$.

It is instructive to examine the physical origin of the dominant
logarithmic correction  proportional to $m^2 \ln | m |$ in 
Eq.~(\ref{eq:gm2}). This term
arises from  a product of two backscattering contributions to the polarization
in Eq.~(\ref{eq:Gibbs2}), which for small $m$ contain also a component involving
only small momentum transfers. 
In fact, if we are only interested in the non-analytic contributions 
to $\Gamma_2 ( N , M )$, we may replace Eq.~(\ref{eq:Gibbs2}) by the simpler
expression
 \begin{eqnarray}
 \Gamma_2^{\rm sing } ( N , M ) & = & 
 - N_L U^2 a^3 \int_{ \tau_{0} }^{\tau_m} d \tau \int_{ - \infty}^{\infty} dx  
 \nonumber
 \\
 & & \hspace{-20mm} \times
 \left[
 \Pi^{ 2 k_{\uparrow} }_{\uparrow} ( x  , \tau )
 \Pi^{ - 2 k_{\downarrow} }_{\downarrow} ( x  , \tau )
+ \Pi^{ - 2 k_{\uparrow} }_{\uparrow} ( x  , \tau )
 \Pi^{  2 k_{\downarrow} }_{\downarrow} ( x  , \tau )
 \right]
 \nonumber
  \\
 &  & \hspace{-24mm} = 
 - N_L U^2 a^3 \frac{2}{ (2 \pi )^4 }
\int_{ \tau_{0} }^{\tau_m} d \tau \int_{ - \infty}^{\infty} dx
\frac{ \cos [ 2 ( k_{\uparrow} - k_{\downarrow} ) x ]  }{ [  x^2 + (v_F \tau)^2 ]^2 }
 \; , 
\nonumber
 \\
 & &
 \label{eq:Gammasing}
 \end{eqnarray}
where $\tau_0 =  ( v_F \Lambda )^{-1}$ and
$\tau_m = ( 4 \pi |m| v_F / a )^{-1}$.  Using
$ k_{\uparrow} - k_{\downarrow} = 2 \pi m /a$ and expanding under the integral sign
for small  $m$
 \begin{equation}
 \cos [ 4 \pi m x /a ] = 1 - \frac{ ( 4 \pi x )^2}{2a^2}  m^2  + O ( m^4 )
 \; ,
 \end{equation}
it is easy to see from Eq.~(\ref{eq:Gammasing}) by power counting
that the coefficient of $m^2$ is decorated by  a non-analytic correction
proportional to $\ln ( \tau_m / \tau_0 ) = \ln  ( a \Lambda /   4 \pi |m| )$.

\section{Discussion and Conclusions}

In this work we have  evaluated the
magnetization-dependent part $g ( m)$ of the 
Gibbs potential $\Gamma ( N , M )$ to
second order in the relevant dimensionless interaction strength 
$I= U \nu_0 $. Our main result is 
an explicit expression for  
$g ( m)$ up to order $I^2$ given in Eq.~(\ref{eq:gm2}), which contains
non-analytic corrections proportional to
$m^2 \ln |m |$ and $  | m|^3$.
 When extrapolated to interactions $I$ of the order of unity, 
these corrections imply that the paramagnetic-ferromagnetic
quantum phase transition in $1D$, if it exists, must be first order.  
Of course, the extrapolation of  the weak coupling expansion 
to values of
$I$ of the order of unity is an uncontrolled procedure, so that
with our method we cannot proof  the existence or the absence
of a ferromagnetic ground state in $1D$. However, the numerical density-matrix renormalization
group calculations by Daul and Noack \cite{Daul98} suggest that in a certain regime
of $\gamma = 4 t^{\prime} /t$ and fillings $n$ 
a ferromagnetic ground state indeed exists.  
On the other hand, these authors found  numerical evidence
that the paramagnetic-ferromagnetic
quantum phase transition in $1D$ is second order, 
which is not supported by our calculation.

Given the fact that the metallic state in $1D$ is a Luttinger liquid,
one could have expected that 
the perturbative expansion of the Gibbs potential
contains non-analytic corrections.
Surprisingly, 
according to a recent calculation by
Betouras, Efremov, and Chubukov \cite{Betouras05}
similar non-analytic corrections  exist even  in higher  dimensions.
These authors evaluated the 
magnetic-field dependent part of the
interaction-correction to  the grand canonical potential  to  second order  
in $U$ in three and two  dimensions.
They found that the
susceptibility is proportional to
 $ h^2  \ln (|h| )$  in 
three dimensions, and proportional to $| h |$ in two dimensions.
These corrections imply non-analyticities in the corresponding 
magnetization-dependent part of the Gibbs potential, leading to 
the breakdown of the Hartree-Fock approximation  in dimensions 
$D > 1$, in agreement
with the work of Belitz and Kirkpatrick \cite{Belitz02}.
 
 Our result (\ref{eq:gm2}) for the Gibbs potential
implies that the magnetic susceptibility at weak coupling
vanishes for $m \rightarrow 0 $  as
 \begin{equation}
 \chi ( m )  = 2 \nu_0 \left[ \frac{\partial^2 g ( m )}{ \partial m^2 } \right]^{-1}
  \sim  \frac{2 \nu_0 }{I^2 \ln ( \lambda / | m | )  }
 \label{eq:chires}
 \; .
 \end{equation} 
Hence, in the paramagnetic state the zero-field susceptibility 
$\chi ( m =0)$ vanishes for any finite value of the interaction,
while it approaches the finite value $ \nu_0$ in the non-interacting limit.
In one-dimensional metals other quantities  are known to exhibit
a similar discontinuity. 
For example, the 
density of states of a Luttinger liquid vanishes at the Fermi energy,
while in the absence of interactions it is finite \cite{Schoenhammer03}. 
We conjecture that higher order corrections neglected in
Eq.~(\ref{eq:gm2}) will transform the logarithmic singularity in Eq.~(\ref{eq:chires})
into a power law with interaction-dependent exponent.

It is important to emphasize that a ferromagnetic ground state in a $1D$ lattice model
with hopping beyond the nearest neighbors does not contradict the
Lieb-Mattis theorem \cite{Lieb62}. Ferromagnetic ground states of the
$t-t^{\prime}$ Hubbard model can therefore not be ruled out a priori.
Whether such a ferromagnetic ground state 
is relevant to explain recent measurements 
of the conductance anomaly in quantum wires \cite{Thomas96,Reilly02} 
is not clear at this point.  
Very recently Klironomos {\it{et al.}} \cite{Klironomos05} 
pointed out that electron-electron interactions in quantum wires
induce deviations from the strictly one-dimensional geometry, in which case
the Lieb-Mattis theorem does not apply and a ferromagnetic ground state
is in principle possible.

A detailed evaluation of the second order correction to the Gibbs potential
in the regime where the Fermi surface without interaction consists of
four points still remains to be done. This calculation  is more
complicated than in the case of two Fermi points, because a self-consistent
treatment within renormalized perturbation
theory requires the introduction of four counterterms, one for each Fermi 
point \cite{Kopietz01,Ledowski05}.  Vollhardt {\it{et al.}} \cite{Vollhardt01}
pointed out that a large asymmetry in the DOS with a peak 
at the lower band edge  tends to favor
ferromagnetism by minimizing the increase of kinetic energy due to
the spin polarization. Given the fact that in our model for $ | \gamma | >1$ the DOS 
exhibits a one-sided singularity at the critical filling $n_c ( \gamma )$ where
the number of Fermi points changes discontinuously 
(see Eqs.~(\ref{eq:ncdef}, \ref{eq:ncdef2}) and Fig.~\ref{fig:dos}), we 
suspect that
for fillings $n$ close to $n_c$ a ferromagnetic ground state can be stabilized even for  
values of the interaction strength $I$ that are substantially smaller than unity. 
Whether this hypothesis is correct
requires a detailed calculation, which is beyond the scope of this work.

\section*{ACKNOWLEDGMENTS}
This work was supported by the DFG via Forschergruppe FOR 412.
We thank Lorenz Bartosch, Nils Hasselmann, Sascha Ledowski, 
Peyman Pirooznia, Francesca Sauli, and
Florian Sch\"{u}tz for  discussions.
One of us (P. Z.) has profited 
from  lectures on the Hubbard model by 
Marcus Kollar at early stages of this work.


\vspace{-0.5mm}
\begin{thebibliography}{99}
\vspace{-0.5mm}
%
\bibitem{Lieb62}
E.\ H.\ Lieb and D.\ Mattis, Phys.\ Rev.\ {\bf{125}}, 164 (1962).
%
\bibitem{Daul98}
S.\ Daul and R.\ M.\ Noack, Phys.\ Rev.\ B {\bf{58}}, 2635 (1998).
%
\bibitem{Bartosch03}
L. Bartosch, M. Kollar, and P. Kopietz, Phys. Rev. B {\bf{67}}, 092403 (2003).
%
\bibitem{Yang04}
K. Yang, Phys. Rev. Lett. {\bf{93}}, 066401 (2004).
%
\bibitem{Sengupta05}
K. Sengupta and Y. B. Kim, Phys. Rev. B {\bf{71}}, 174427 (2005).
%
\bibitem{Belitz02}
D. Belitz and T. R. Kirkpatrick, 
Phys. Rev. Lett. {\bf{89}}, 247202 (2002).
%
\bibitem{Nozieres86}
P.\ Nozi\`{e}res,  {\it{Ferromagnetisme Itin\'{e}rant}} (Lecture Notes, 1986, unpublished).
%
\bibitem{Georges91}
A. Georges and J. S. Yedidia, Phys. Rev B {\bf{43}}, 3475 (1991).
%
\bibitem{Kollar03}
M. Kollar, I. Spremo, and P. Kopietz,
Phys. Rev. B {\bf{67}}, 104427 (2003).
%
\bibitem{Moriya85}
T.\ Moriya, {\it{Spin Fluctuations in Itinerant Electron Magnetism}}
(Springer, Berlin, 1985).
%
\bibitem{Nozieres64}
P. Nozi\`{e}res, {\it{Theory of Interacting Fermi Systems}}, (Benjamin,
New York, 1964).
%
\bibitem{footnote1}
In the regime where the Fermi surface without
symmetry breaking and magnetic field consists of four points,
the number of counterterms must also be doubled.
We shall not further discuss this regime in this work.
%
\bibitem{Neumayr03}
A. Neumayr and W. Metzner, Phys. Rev. B {\bf{67}}, 035112 (2003).
%
\bibitem{Kopietz01}
P. Kopietz and T. Busche, Phys. Rev. B {\bf{64}}, 155101 (2001).
%
\bibitem{Ledowski05}
S. Ledowski, P. Kopietz, and A. Ferraz, Phys. Rev. B {\bf{71}}, 235106 (2005).
%
\bibitem{Betouras05}
J. Betouras, D. Efremov, and A. Chubukov,
cond-mat/0506083.
%
\bibitem{Schoenhammer03}
See, for example, K. Sch\"{o}nhammer, in {\it{Strong Interactions
in Low Dimensions}}, edited by D. Baeriswyl and L. Degiorgi
(Dordrecht, Kluwer, 2003).
%
\bibitem{Thomas96}
K.\ J.\ Thomas, J.\ T.\ Nicholls, M.\ Y.\ Simmons, M.\ Pepper, D.\ R.\ Mace, and D.\ A.\ Ritchie, Phys.\ Rev.\ Lett.\ {\bf{77}}, 
135 (1996).
%
\bibitem{Reilly02}
D.\ J.\ Reilly,
 T.\ M.\ Buehler, J.\ L.\ O'Brien, A.\ R.\ Hamilton, 
A.\ S.\ Dzurak, R.\ G.\ Clark, B.\ E.\ Kane, L.\ N.\ Pfeiffer, and K.\ W.\ West,
Phys. Rev. Lett. {\bf{89}}, 246801 (2002).
%
\bibitem{Klironomos05}
A. D. Klironomos, J. S. Meyer, and K. A. Matveev, cond-mat/0507387.
%
\bibitem{Vollhardt01}
D.\ Vollhardt, N.\ Bl\"{u}mer, K.\ Held, and  M.\ Kollar,
in {\it{Band Ferromagnetism}}, eds.\ K.\ Baberschke, M.\ Donath, and
W.\ Nolting (Springer, Heidelberg, 2001), p.\ 191.
%

\end{thebibliography}
\end{document}